\documentstyle[aps,preprint,eqsecnum,epsfig,tighten]{revtex}
\input epsf.tex

\begin{document}

\draft

\title{\bf PION-PION SCATTERING IN TWO DIMENSIONS}
\author{D. Delphenich\footnote{delpheni@suhep.syr.edu},
        J. Schechter \footnote{schechte@@suhep.syr.edu}
         and S. Vaidya \footnote{sachin@suhep.phy.syr.edu}
}
\address{\em Department of Physics, Syracuse University,
             \\ Syracuse, N. Y.  13244-1130,  U. S. A. 
        \\
}
\preprint{\vbox{\hbox{SU-4240-679} \hbox{June 1998}}}

\maketitle

\begin{abstract}
Massive two-flavor $QED_2$ is known to have many similarities to the
two-flavor $QCD_4$. Here we compare the $\pi-\pi$ scattering amplitudes
(actually an analog process in $QED_2$) of the two theories. The $QED_2$
amplitude is computed from the bosonized version of the model while the
$QCD_4$ amplitude is computed from an effective low energy chiral
Lagrangian. A number of interesting features are noted. For example, the
contribution of the two-dimensional Wess-Zumino-Witten (WZW) term in
$QED_2$ is structurally identical to the vector meson exchange contribution
in $QCD_4$. Also it is shown that the $QED_2$ amplitude computed at
tree level is a reasonable approximation to the known exact strong coupling
solution.
\end{abstract}

\pacs{11.10.Kk, 11.30.Rd, 13.75.Lb}

\section{Introduction}
Two dimensional massive QED ($QED_2$)
\cite{schwinger1,schwinger2,lowswi,coleman2,gepner,efhk,shismi,hehois} and
four dimensional QCD ($QCD_4$) both are asymptotically free, confining
theories with non-trivial vacuums labeled by an angular parameter
$\theta$. This similarity has been exploited for theoretical purposes in a
great body of work, nicely summarized in \cite{Ababro,frison} for
example. The similarity is enhanced if one examines the non-Abelian
bosonized \cite{witten4,novikov} version \cite{gepner,delsch1} of massive
multi-flavor $QED_2$. In this version the action is given by
Eqs. (\ref{action}) and (\ref{wzwterm}) below together with the additional
piece
\begin{equation}
-\frac{e^2}{8 \pi^2} \int (\theta + i\;\text{ln}\;\text{det} \;U)^2 d^2x,
\end{equation}
which decouples as the coupling strength $e$ gets large. Now it will not
escape the attentive reader that this looks identical to the low energy
{\it effective} action made from pion fields for ordinary $QCD_4$, if only
allowance is made for the two extra dimensions which must appear.
Specifically, the non-linear sigma model terms in the form (\ref{action})
look like those in, for instance, \cite{cronin}, the term (\ref{wzwterm})
looks like the one in \cite{witten3} and the term involving $\theta$ above
looks like the one in
\cite{rosctr,divven,witten2,natarn,autato,kawoht}. Since some exact results
are known for $QED_2$ it is reasonable to expect to learn something new
about the effective Lagrangian approach to $QCD_4$. At least, it is
interesting to test the accuracy of the tree approximation in $QED_2$; that
is not so easy to do in $QCD_4$ where the exact analytic results are not
known.

In an earlier paper \cite{delsch1}, among other things, the two-point
functions of the two theories were compared. This led to a simple
understanding, at the tree level, of the stability of the equality
$m(\pi^+)=m(\pi^0)$ to addition of a minimal isotopic spin violating term
in the strong coupling limit of two flavor massive $QED_2$. This was seen
to be the analog of the fact that in ordinary QCD, the $\pi^+ - \pi^0$ mass
difference is essentially unrelated to the up-down quark mass
difference. In the present paper we will carry out the tree level
comparison of the two theories for their four-point functions, i.e. the
$\pi-\pi$ scattering amplitudes. It will be seen that a number of
surprising features emerge. These features are related to the not so
innocuous fact that the two dimensional analog pion field has {\it
positive} $G$-parity, unlike the four dimensional case. This has the
consequence that a three pion vertex is allowed in two dimensions and
appears in the ``topological'' Wess-Zumino-Witten \cite{witten4,novikov}
term. The form of the scattering amplitude which results from pion exchange
is not similar to anything in the four dimensional scattering amplitude
computed from an effective chiral Lagrangian of only pseudoscalars. Rather,
it is identical in structure to the {\it vector} meson exchange graphs in
the four dimensional model based on a chiral Lagrangian of pseudoscsalars
{\it plus} vectors. It has been known for many years that the addition of
vector mesons to the pion effective Lagrangian substantially improves the
tree level predictions. It is intriguing that an exact bosonized analog
model leads to just this type of structure.

We will specifically focus attention on massive, isospin invariant
two-flavor $QED_2$ \cite{coleman2} in the strong coupling
approximation. The stable light particle spectrum consists of an analog
pion (pseudoscalar, isotriplet) and an analog sigma meson (scalar
isosinglet). The theory will be treated in the bosonized format. There are
two possible ways to bosonize. In the Abelian method \cite{coleman2}, the
Lagrangian is constructed only from the $\pi^0$ field. The $\pi^+$ and
$\pi^-$ states appear as ``static'' solitons while the $\pi^0$ appears
again as a time-dependent ``breather'' soliton. The $\sigma$ appears as a
second breather soliton. In the non-Abelian method, the Lagrangian is
constructed from the complete $\pi$ triplet. The triplet appears again as a
static soliton and still\footnote{This is the interpretation of
\cite{delsch1}. A slightly different interpretation is given in
\cite{gepner}.} once more as the first breather. The $\sigma$ appears as
the second breather.

We shall work here with the tree level non-Abelian bosonized action, which
has the advantage of manifest isotopic spin invariance. On the other hand,
for the consideration of the solitons of the model, the Abelian bosonized
version is much easier to work with since there is less redundancy. Note
that the $\sigma$ state only appears non-perturbatively (as a soliton). We
can, as is done in four dimensional effective theories \cite{ccwz}, add it
in an {\it ad hoc} way at the tree level, remembering that a more exact
treatment of the model would render the tree level term unnecessary. The
question of a $\sigma$-meson in the four dimensional effective QCD
Lagrangian is very timely inasmuch as a number of authors 
\cite{hasasc,torroo,iititt,morpen,jphs,bmvp,andman,vereshagin,verver,achshe,kalema,svec,brmdrr,delsca,athasa,olospe}
have recently provided evidence that such a state ought to exist.

Section 2 contains the set up of the model and the computation of the tree
level analog $\pi-\pi$ scattering amplitudes for arbitrary particle
charges. These are displayed using the redundant set (more so in two
dimensions than in four dimensions) of Mandelstam variables. Comparison is
made with four-dimensional $\pi-\pi$ scattering and some interesting
features are noted. In Section 3 we mainly focus on the specific case of
the analog $\pi^+ - \pi^-$ scattering and eliminate the redundancy of the
Mandelstam variable description by specializing to the transmission and
reflection amplitudes. The exact solution in the strong coupling limit,
based on known results \cite{zamzam,rajaraman} for the sine-Gordon theory,
is written in Section 4. It is observed that this limit corresponds to pure
reflectionless scattering. However, the theory is not trivial because it
contains two bound state poles. The tree level amplitude is shown to give
an accurate, approximate value for the residue at the analog pion
pole. Finally, Section 5 contains a brief summary, discussion of the
significance of these results and some directions for future work.

\section{Scattering in the Tree Level Non-Abelian Bosonized Model}

Many authors have observed that the ordinary low energy $3+1$ dimensional
QCD has a number of striking similarities to $1+1$ dimensional two-flavor
QED \cite{Ababro,frison}. Now, in the ordinary QCD case, the scattering of
pions is already quite well described near threshold by the tree level
treatment of the effective Lagrangian of pions (Of course, further higher
order improvements can be implemented in the chiral perturbation scheme
\cite{weinberg,gasleu1,gasleu2,meissner}). It is thus interesting to
investigate the tree level scattering of the analog ``pions'' in the
two-dimensional, two-flavor QED where, as we will discuss, an exact answer
is available for comparison. This may be a useful step in obtaining a
deeper understanding of four-dimensional QCD and its relation to the
two-dimensional model. In fact, we shall see an initially unexpected
correspondence between the two theories. Naturally there are important
differences as well. One question of interest concerns the low mass
$\sigma$-meson and its two-dimensional analog. There has been a great deal
of recent work
\cite{hasasc,torroo,iititt,morpen,jphs,bmvp,andman,vereshagin,verver,achshe,kalema,svec,brmdrr,delsca,athasa,olospe}
on the possibility of an experimental $\sigma$-meson in QCD. At the same
time the two-dimensional model is known \cite{coleman2} to contain an
analog $\sigma$-meson with mass $m(\sigma) \approx \sqrt{3} m(\pi)$; this
is however a bound state rather than a scattering resonance.

In this section we shall calculate the tree-level analog pion-pion
scattering $S$-matrix starting from the bosonized two-flavor QED
Lagrangian. Since this Lagrangian only contains ``pion'' fields we shall
also consider adding a suitable extra piece to get the $\sigma$-meson pole
at the tree level.

The relevant non-Abelian bosonization method was developed by Witten
\cite{witten4} and first applied to two-flavor $QED_2$ by Gepner
\cite{gepner}. Here we shall follow the notation of \cite{delsch1}; the
bosonized action is given in (2.10) of this reference. In the strong
coupling limit, an $\eta$-type particle (pseudoscalar isosinglet) decouples
and the effective action is given in terms of a unitary unimodular matrix
field $U(x)$ which transforms under left and right chiral transformations
$U_L, U_R$, as $U(x) \rightarrow U_L U(x) U^{\dagger}_R$ and which has the
decomposition
\begin{equation}
U(x) = e^{i \sqrt{4 \pi} \phi(x)}, \quad \phi(x) =
\frac{1}{\sqrt{2}}\vec{\tau}. \vec{\pi}(x).
\label{uandphi}
\end{equation}

Here the $\tau_i$ are the Pauli matrices and the $\pi_i(x)$ are the analog
pion fields. Actually all the $\pi_i$ carry zero electric charge in two
flavor $QED_2$ but for convenience we will assign the names
\begin{equation}
\pi^{\pm} = \frac{1}{\sqrt{2}} (\pi_1 \mp i \pi_2), \quad \pi^0 = \pi_3.
\end{equation}
The low energy bosonized action of multi-flavor $QED_2$ reads
\begin{equation}
\Gamma = \int d^2x [-\frac{1}{8 \pi} Tr (\partial_\mu U \partial_\mu
U^{\dagger}) + \frac{m^2}{2} Tr (U + U^{\dagger} -2)] + \Gamma_{WZW}
\label{action}
\end{equation}
where the third (Wess-Zumino-Witten) term \cite{witten4,novikov} may be
compactly written using the matrix one-form $\alpha = dU U^{\dagger}$ as
\begin{equation}
\Gamma_{\text{WZW}} = \frac{1}{12 \pi} \int_{M^3} \text{Tr}\; (\alpha^3).
\label{wzwterm}
\end{equation}
Here, $M^3$ is a three-dimensional manifold whose boundary $\partial M^3$
is the two-dimensional Minkowski space. Note that as written,
(\ref{action}) and (\ref{wzwterm}) can be used for an arbitrary number
$N_f$ of flavors; we are however specializing to the case $N_f = 2$ by
restricting the matrix $U$ to be a $2 \times 2$ matrix in form. Furthermore
we have, unlike \cite{delsch1}, restricted $U$ to satisfy $det U = 1$ as is
appropriate for low energies where the pseudoscalar isosinglet may be
considered infinitely heavy.

Now for a perturbative treatment of $\pi-\pi$ scattering we should expand
(\ref{action}) up to fourth order in the number of pion fields. The
quadratic terms arise from the first two terms of (\ref{action}) and give
the Lagrangian 
\begin{equation}
{\cal L}^{(2)} = -\frac{1}{2} \partial_\mu \vec{\pi}. \partial_\mu \vec{\pi}
- \frac{m_{\pi}^2}{2} \vec{\pi}. \vec{\pi},
\end{equation}
wherein we have identified 
\begin{equation}
m_{\pi} = 2 \sqrt{\pi} m.
\end{equation}
We have used Eq. (\ref{uandphi}) in obtaining this result. The first two
terms of (\ref{action}) also yield quartic terms which may be simplified
to 
\begin{equation}
{\cal L}^{(4)} = \frac{\pi}{3}[(\partial_\mu \vec{\pi}. \partial_\mu
\vec{\pi})(\vec{\pi}. \vec{\pi}) - (\vec{\pi}. \partial_\mu \vec{\pi})^2] +
\frac{\pi}{12} m_{\pi}^2 (\vec{\pi}. \vec{\pi})^2.
\label{contactterm}
\end{equation}
Finally the Wess-Zumino-Witten term yields a cubic interaction of the pion
fields;
\begin{equation}
{\cal L}^{(3)} = \frac{i \sqrt{2 \pi}}{3} \epsilon_{\mu \nu}\epsilon_{jkl}
\pi_j \partial_\mu \pi_k \partial_\nu \pi_l
\label{cubicterm}
\end{equation}
where $\epsilon_{12} = -\epsilon_{21} = 1$. In obtaining (\ref{cubicterm}),
we used Stokes' theorem as
\begin{equation}
\int_{M^3} \text{Tr}(d \phi\; d \phi\; d \phi) = \int_{M^3}d \text{Tr}(\phi
\;d \phi \; d \phi) = \int_{\partial M^3} \text{Tr}( \phi\; d \phi\; d \phi).
\end{equation}

Judging from usual experience with four dimensional physics,
Eq.~(\ref{cubicterm}) may appear startling. While it may seem to be ruled
out because the two-dimensional pions are pseudoscalar, the $\epsilon_{\mu
\nu}$ factor rescues parity (This will not work in four dimensions.). In
four dimensions, a three-pion vertex is also ruled out because the pion
carries a negative $G$-parity. In two dimensions however a pseudoscalar
bilinear $\bar{\psi} \gamma_5 \psi$ picks up a minus sign under charge
conjugation so, taking the isotopic spin factor in $G=e^{i \pi I_2}C$ into
account, the two-dimensional analog pion carries {\it positive} $G$-parity.

The momentum space trilinear pion interaction (essentially $i {\cal
L}^{(3)}$ from (\ref{cubicterm})) or ``Feynman rule'' is:
\begin{equation}
2 \sqrt{2 \pi} \epsilon_{ijk} \epsilon_{\mu \nu} p^{(j)}_\mu p^{(k)}_\nu.
\label{cubicfr}
\end{equation}

\begin{figure}[h]
\centerline{\epsfig{file=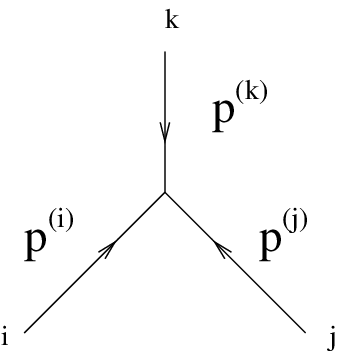,clip=2in,width=2in}}
\caption{Feynman diagram for the cubic pion interaction. For example
$p^{(i)}$ is the 2-momentum of the pion with isotopic spin index $i$. Also
$p^{(i)} + p^{(j)} + p^{(k)} = 0$.}
\label{fig:fd}
\end{figure}

\noindent This corresponds to the diagram in Fig. \ref{fig:fd}. 

We are interested in the analog pion-pion scattering reaction:
\begin{equation}
\pi_i(p_1) + \pi_j(p_2) \rightarrow \pi_k(p'_1) + \pi_l(p'_2),
\label{pipireaction}
\end{equation}
where $i, j, k, l$ are the isospin indices and the $(p_i)_\mu$ are the
two-momenta. The usual Mandelstam variables are
\begin{eqnarray}
s=-(p_1 + p_2)^2, \; t&=&-(p_1 - p'_1)^2, \; u=-(p_1 - p'_2)^2,
\label{mandelstam}\\ 
s+t+u &=& 4m_{\pi}^2.
\end{eqnarray}
In the present $1+1$ dimensional case, these are somewhat redundant as the
only two physical possibilities are 
(a) forward scattering in the center-of-mass frame:
\begin{equation}
t=0, \quad u=4m_{\pi}^2 -s,
\label{forward}
\end{equation}
and
(b) backward scattering in the center-of-mass frame:
\begin{equation}
u=0, \quad t= 4m_{\pi}^2 -s,
\label{backward}
\end{equation}
With the one-particle state normalization
\begin{equation}
<\pi_j(p')|\pi_i(p)> = \delta_{ij} \delta(p - p'),
\end{equation}
the standard crossing-symmetric parameterization of the scattering matrix
element for the reaction (\ref{pipireaction}) is 
\begin{eqnarray}
\lefteqn{<\pi_k(p'_1) \pi_l(p'_2)|S|\pi_i(p_1)\pi_j(p_2)> =} \nonumber \\
& & \delta_{ik}\delta_{jl} \delta(p_1 - p'_1)\delta(p_2 - p'_2) +
 \delta_{il}\delta_{jk} \delta(p_1 - p'_2)\delta(p_2 - p'_1) + \nonumber \\
& &\frac{i}{4} \frac{1}{\sqrt{E_1 E_2 E'_1 E'_2}}\delta(p_1+p_2-p'_1-p'_2)
 \delta(E_1+E_2-E'_1-E'_2) \times \nonumber \\
&\times& \Big[ \delta_{ij}\delta_{kl}A(s,t,u) +
\delta_{ik}\delta_{jl}A(t,s,u) + \delta_{il}\delta_{jk}A(u,t,s) \Big] 
\label{smatrix}
\end{eqnarray}
This is a convenient form for perturbation theory and for comparison with
the four-dimensional case. All the dynamics is contained in the function
$A(s, t, u)$. 

Our tree-level perturbation calculation yields
\begin{equation}
A(s,t,u)= 2\pi(s-m_{\pi}^2) + 2\pi \left[ \frac{t(s-u)}{m_{\pi}^2-t} +
\frac{u(s-t)}{m_{\pi}^2 - u} \right] + \gamma^2 \frac{(s-2m_{\pi}^2)^2} 
{m_{\sigma}^2 - s}.
\label{samplitude}
\end{equation}
The first term arises from the contact interaction (\ref{contactterm})
while the second term is associated with pion exchange diagrams using the
vertices from (\ref{cubicterm}) or, perhaps more conveniently, from
(\ref{cubicfr}). The third term does not follow from the tree level
treatment of the action (\ref{action}) but was added on somewhat ad hoc
grounds for comparison with the four-dimensional case. We expect, as
discussed in Section 1, that a $\sigma$-particle (and hence a $\sigma$
pole in the scattering amplitude) should arise as a ``breather soliton''
from (\ref{action}). We may formally treat the $\sigma$ as a ``matter''
particle according to the method of \cite{ccwz}; then the third term in
(\ref{samplitude}) comes from the following addition to the interaction
Lagrangian:
\begin{equation}
{\cal L}_{\sigma \pi \pi} = \frac{\gamma}{4 \pi} \sigma \; \text{Tr}\; 
(\partial_\mu U \partial_\mu U^{\dagger}),
\end{equation}
where $\gamma$ is a real coupling constant.

It is interesting to compare the two-dimensional $\pi - \pi$ scattering
amplitude (\ref{samplitude}) with a recently considered model \cite{hasasc}
which gives a reasonable phenomenological description of ordinary pion
scattering (presumably four-dimensional QCD) up to 1.2 GeV. That model,
prompted by the $1/N_c$ expansion \cite{witten1,hooft}, starts out by
writing the amplitude as the tree expansion of a chiral Lagrangian
including scalar mesons. The model is formally crossing symmetric but, for
arbitrary choice of parameters, may very badly violate unitarity bounds. A
kind of ``regularization'' in the vicinity of the physical divergences at
the direct channel poles is performed which formally maintains crossing
symmetry. Then the arbitrary parameters are adjusted to provide
cancellations which preserve the unitarity bounds (and fit the data). In
this way an approximate amplitude obeying both crossing symmetry and
unitarity is obtained. The unregularized amplitude (see Eqs. (C1), (C2),
and (C3) of \cite{hasasc}) for this model is
\begin{eqnarray}
A^{QCD}(s,t,u) &=& \frac{2(s-m_{\pi}^2)}{F_{\pi}^2} + \frac{g_{\rho \pi
\pi}^2}{2 m_{\rho}^2} \left[ \frac{t(s-u)}{m_{\rho}^2 -t} +
\frac{u(s-t)}{m_{\rho}^2 - u} \right] \nonumber \\
&+& \frac{\gamma_0^2}{2} \frac{(s-2m_{\pi}^2)^2}{m_{\sigma}^2-s} + \dots 
\label{qcdamplitude}
\end{eqnarray}
where $F_{\pi} \approx 0.131$ GeV is the pion decay constant, $m_{\rho}$ is
the mass of the $\rho$-meson and $g_{\rho \pi \pi}$ is the $\rho \pi \pi$
coupling constant. The 3 dots stand for another scalar meson pole term
which however is not expected to exist in the two-flavor version of the
model. The first term of (\ref{qcdamplitude}) is the ``current algebra''
term which is well-known to be a good approximation very close to the $\pi
\pi$ threshold. It is not very surprising that it is identical, up to a
numerical factor, to the first term of (\ref{samplitude}). Similarly it is
not surprising that the third, $\sigma$-meson exchange term in
(\ref{qcdamplitude}), is identical up to a factor with the third term in
(\ref{samplitude}). What is much more surprising is that the second term in
(\ref{qcdamplitude}), which represents the effects of the chiral-symmetric
$\rho$-meson exchange, has the same structure (up to an overall numerical
factor and with the replacement $m_{\rho} \rightarrow m_{\pi}$) as the
second (WZW) term in (\ref{samplitude}). From a technical standpoint it may
be reasonable in the sense that both the rho and the pion are isovector
particles. However it is amusing to see that the analog of the
two-dimensional WZW model of ``pions'' is not the four dimensional WZW
model of pions but must also include the terms associated with the
introduction of the $\rho$-meson in a chiral-symmetric manner.

Let us now try to exploit the above correspondence. We notice that the {\it
first terms} of (\ref{samplitude}) and (\ref{qcdamplitude}) satisfy
\begin{equation}
A(s,t,u)= \pi F^2_{\pi} A^{QCD}(s,t,u).
\label{firstequal}
\end{equation}
Suppose we assume that the corresponding second terms also obey
(\ref{firstequal}) in the $m_{\rho}=m_{\pi}$ limit. This then demands that 
\begin{equation}
\frac{F_{\pi}^2 g_{\rho \pi \pi}^2}{2 m_{\rho}^2} =2,
\label{krsf}
\end{equation}
which is of the form of the famous KSRF \cite{kawsuz,riafay} relation (which, however, has 1 rather than 2 on the right hand side). From the present perspective, this
KSRF-type relation is the analog of the special relationship which exists
between the kinetic (first term of (\ref{action})) and the ``topological''
(\ref{wzwterm}) term of the two-dimensional WZW model. In the two
dimensional model, this special relationship between the kinetic and the
topological terms is required \cite{witten4} to obtain the correct
equations of motion and currents.

\section{Formulas for Particular Reactions}
In the previous section we discussed the formal analogy between the
two-flavor ${QED}_2$ and ${QCD}_4$ scattering amplitudes at tree level. Now
let us concentrate on the two-dimensional scattering itself in more
detail. We will consider various ``charged'' meson reactions with
appropriate two-dimensional kinematics. Eq.(\ref{smatrix}) contains
information about the scattering of pions with all ``charges''. It is
standard (see for example p 178 of \cite{DGH}) to consider linear
combinations $T^{(I)}(s,t,u)$ corresponding to scattering states of
definite isotopic spin $I=0,1,2$:
\begin{eqnarray}
T^{(0)}(s,t,u) &=& 3 A(s,t,u) + A(t,s,u) + A(u,t,s), \\
T^{(1)}(s,t,u) &=& A(t,s,u) - A(u,t,s), \\
T^{(2)}(s,t,u) &=& A(t,s,u) + A(u,t,s).
\end{eqnarray}
Amplitudes for scattering ``pions'' with definite ``charges'' are related
to these; for example
\begin{eqnarray}
T^{(+-)} &=& \frac{1}{3}T^{(0)} + \frac{1}{2}T^{(1)} + \frac{1}{6}T^{(2)} =
A(s,t,u) + A(t,s,u), \label{pi+pi-} \\
T^{(+0)} &=& \frac{1}{2}T^{(1)} + \frac{1}{2}T^{(2)} = A(t,s,u), \\
T^{(++)} &=& T^{(2)} = A(t,s,u) + A(u,t,s) \label{pi+pi+},
\end{eqnarray}
etc.

It is also desirable to eliminate the large redundancy in the kinematical
description of two dimensional scattering when the Mandelstam variables
(\ref{mandelstam}) are used. We should specialize to the two cases: forward
and backward scattering in the center of mass frame as specified in
(\ref{forward}) and (\ref{backward}). This may be conveniently implemented
by re-expressing the overall energy-momentum conservation delta function in
(\ref{smatrix}) as 
\begin{eqnarray}
\lefteqn{\delta(p_1 + p_2 - p'_1 - p'_2) \delta(E_1 + E_2 - E'_1 - E'_2)=}
\nonumber \\
& &\left| \frac{p_1}{E_1} -\frac{p_2}{E_2} \right|^{-1} \left[ \delta(p_1 -
p'_1) \delta(p_2 - p'_2) + \delta(p_1 - p'_2)\delta(p_2 - p'_1) \right].
\label{emdelta}
\end{eqnarray}
This can be verified by multiplying both sides by an arbitrary ``test
function'' and integrating. The first term on the right hand side of
(\ref{emdelta}) enforces a forward scattering evaluation while the second
term yields the backward scattering evaluation. A needed factor in
(\ref{smatrix}) is evaluated as 
\begin{equation}
\displaystyle{\frac{\left| \frac{p_1}{E_1} - \frac{p_2}{E_2}
\right|^{-1}}{\sqrt{E_1 E_2 E'_1 E'_2}}} =
\frac{2}{\sqrt{s(s-4m_{\pi}^2)}}.
\label{fluxfactor}
\end{equation}
We will mainly be interested in the $\pi^+ \pi^- \rightarrow \pi^+ \pi^-$
reaction. According to (\ref{pi+pi-}) it corresponds to the linear
combination $A(s,t,u)+A(t,s,u)$. The S-matrix, after using (\ref{emdelta}),
can be written as the sum of a ``transmission'' piece (proportional to
$\delta(p_1 - p'_1) \delta(p_2 - p'_2)$) and a ``reflection'' piece
proportional to $\delta(p_1 - p'_2) \delta(p_2 - p'_1)$:
\begin{equation}
S^{(+-)} = \delta(p_1 - p'_1) \delta(p_2 - p'_2)S^{(+-)}_T + \delta(p_1 -
p'_2)\delta(p_2 - p'_1)S^{(+-)}_R. 
\end{equation}
With the help of (\ref{forward}), (\ref{backward}) and (\ref{fluxfactor})
we find from (\ref{smatrix}) and (\ref{samplitude}):
\begin{eqnarray}
S^{(+-)}_T &=& 1+ \frac{i}{2 \sqrt{s(s-4m_{\pi}^2)}} \left[
2\pi(s-2m_{\pi}^2)+2\pi \frac{s(s-4m_{\pi}^2)}{m_{\pi}^2-s} \right.\nonumber \\
&& + \left. \gamma^2 \left( \frac{(s-2m_{\pi}^2)^2}{m_{\sigma}^2-s} +
\frac{4m_{\pi}^4}{m_{\sigma}^2} \right) \right], \label{transmission} \\
S^{(+-)}_R &=& \frac{i}{2 \sqrt{s(s-4m_{\pi}^2)}} \left[
4 \pi m_{\pi}^2 +2\pi s (4m_{\pi}^2-s)
\left(\frac{1}{m_{\pi}^2-s}-\frac{1}{3m_{\pi}^2-s}
\right) \right.\nonumber \\ 
& & + \left. \gamma^2 (s-2m_{\pi}^2)^2 \left( \frac{1}{m_{\sigma}^2-s} -
\frac{1}{(4m_{\pi}^2-m_{\sigma}^2)-s} \right) \right]. \label{reflection}
\end{eqnarray}
The $1$ on the left hand side of (\ref{transmission}) but not
(\ref{reflection}) is due to resolving the unit operator in
(\ref{smatrix}) analogously to the resolution of the amplitude in
(\ref{pi+pi-}). As in ordinary one-dimensional quantum mechanics,
$S^{(+-)}_T \rightarrow 1$ and $S^{(+-)}_R \rightarrow 0$ when the
interaction vanishes. Another check is provided by considering the S-matrix
for $\pi^+-\pi^+$ scattering. In this case Bose statistics prohibits any
distinction between forward and backward scattering. The appropriate linear
combination (see (\ref{pi+pi+})) for $\pi^+ \pi^+ \rightarrow \pi^+ \pi^+$
is $A(t,s,u) + A(u,t,s)$ and it is easy to see that both (\ref{forward})
and (\ref{backward}) give the same results for this quantity. Then
(\ref{smatrix}) becomes
\begin{eqnarray}
\lefteqn{S^{(++)}=[\delta(p_1-p'_1)\delta(p_2-p'_2) +
\delta(p_1-p'_2)\delta(p_2-p'_1)]\times } \nonumber \\
& &\times \left[1+\frac{i}{2 \sqrt{s(s-4m_{\pi}^2)}}
\left(2\pi(2m_{\pi}^2 -s)+2\pi \frac{s(s-4m_{\pi}^2)}{s-3m_{\pi}^2}\right)
\right. \nonumber \\ 
& & + \left. \gamma^2 \left(\frac{(s-2m_{\pi}^2)^2}
{m_{\sigma}^2-4m_{\pi}^2+s} + \frac{4m_{\pi}^4}{m_{\sigma}^2} \right) \right], 
\label{spi+pi+}
\end{eqnarray}
which does not distinguish between forward and backward scattering. 

Let us focus on the transmission amplitude $S^{(+-)}_T$ for
definiteness. Note again that the term proportional to $\gamma^2$ does not
follow from the perturbative treatment of the original action
(\ref{action}). It has been introduced to mimic the $\sigma$ meson exchange
(in analogy to the treatment of 2 flavor $QCD_4$) since it is known that a
$\sigma$ meson with mass satisfying $m_{\sigma}^2 = 3 m_{\pi}^2$ should
exist in the strong coupling limit. Without the $\gamma^2$ term, the model
just differs from the WZW action \cite{witten4} by the mass term. If the
mass term is also dropped we would have just the WZW action which describes
the free theory.  As another check, we observe that $S^{(+-)}_T$ does in
fact go to 1 when $\gamma$ and $m_{\pi}^2$ are set to zero. When
$\gamma=0$, the large $s$ behaviors are
\begin{equation}
S^{(+-)}_T \rightarrow 1 + \frac{i \pi m_{\pi}^2}{s}, \quad S^{(+-)}_R
\rightarrow \frac{i 6 \pi m_{\pi}^6}{s^3}.
\label{larges}
\end{equation}
$S^{(+-)}_T$ has poles at $s=m_{\pi}^2$ and $s=m_{\sigma}^2 = 3
m_{\pi}^2$. Both are below the threshold at $s=4m_{\pi}^2$ and therefore to
be interpreted as bound states. For comparison with the work in the next
section we give the residues:
\begin{eqnarray}
\text{Res}[S^{(+-)}_T, s=m_{\pi}^2] &=& \mp \sqrt{3} \pi m_{\pi}^2,
\label{treeresidue1} \\
\text{Res}[S^{(+-)}_T, s=3m_{\pi}^2]&=& \mp \frac{\gamma^2 m_{\pi}^2} {2
\sqrt{3}}, \label{treeresidue2}
\end{eqnarray}
where the $\mp$ corresponds to the different sign choices for the square
root in (\ref{transmission}).

\section{Connection with Exact Results}
Coleman \cite{coleman2} has argued that two flavor massive $QED_2$ with
isotopic spin invariance reduces simply to the sine-Gordon theory in the
strong coupling limit, when attention is focussed on the light particles of
the theory. This follows from the treatment of the model by the Abelian
bosonization technique. That approach requires two pseudoscalar fields
$\chi_1$ and $\chi_2$ which enter into the Lagrangian as
\begin{equation}
{\cal L}_{\text{Abelian}} = -\frac{1}{2}\sum_{i=1,2} \partial_\mu \chi_i
\partial_\mu \chi_i - \frac{e^2}{2 \pi} (\sum_i \chi_i - \frac{\theta}{2
\pi})^2 - m^2 \sum_i[1-\text{cos}(2 \sqrt{\pi} \chi_i)]
\label{abelianL}
\end{equation}
where $\theta$, (which will be set to zero henceforth), represents the
background electric field of the underlying theory. The Lagrangian
(\ref{abelianL}) has the same form as the one for the classical soliton
ansatz of the non-Abelian bosonized model (see Eq.(4.2) of
\cite{delsch1}). It is convenient to define $\pi^0$ and $\eta$ as 
\begin{equation}
\pi^0 = \frac{\chi_1 - \chi_2}{\sqrt{2}}, \quad \eta = \frac{\chi_1 +
\chi_2}{\sqrt{2}}.
\end{equation}
Notice that both the electric charge $e$ and the mass parameter $m$ have
the same units. When $e \gg m$ (i.e. strong coupling) the $\eta$ field
becomes very heavy and decouples. We are then left with a special case of
the sine-Gordon model:
\begin{equation}
{\cal L}_{\text{Abelian}} \rightarrow -\frac{1}{2}(\partial_\mu \pi^0)^2 +
2 m^2 \; \text{cos}(\sqrt{2 \pi} \pi^0).
\label{strongcoupling}
\end{equation}
This enables us to read off a tree level $\pi^0$ mass of $2\sqrt{\pi}m
\approx 3.54m$. The $\pi^+$ and $\pi^-$ particles are hidden from sight in
(\ref{strongcoupling}) but appear \cite{coleman2} as solitons and
anti-solitons with mass
\begin{equation}
M = \frac{8m}{\sqrt{\pi}} \approx 4.51 m.
\end{equation}
At first glance it appears that the $\pi^{\pm}$ masses differ from the
$\pi^0$ mass. However the theory also contains two ``breather'' solitons
with masses given by \cite{dahane}:
\begin{equation}
M_n = 2M \;\text{sin} \left( \frac{n \pi}{6} \right), \quad n=1,2.
\label{breathermass}
\end{equation}
This formula is argued to be exact when $M$ includes radiative corrections
to the classical soliton mass. The $n=1$ breather has mass $M_1 = M$ and is
identified by Coleman as the $\pi^0$, thereby restoring the isotopic spin
invariance. The tree level pion mass is considered to be just a rough (to
about 20\% accuracy) approximation. Finally the $n=2$ breather is
identified as the isosinglet sigma with mass
$m_{\sigma}=\sqrt{3}m_{\pi}$. 

Finding the exact solution for the scattering matrix in the sine-Gordon
model is a by now classic problem which has been solved and elucidated by
several authors \cite{zamzam}. It is carried out for a more general
sine-Gordon model than (\ref{strongcoupling}); there is an extra parameter
$\gamma$ which shows up by (\ref{breathermass}) now reading $M_n=2M\;
\text{sin}(n \gamma/16)$. Clearly we are dealing with the special case
\begin{equation}
\gamma=\frac{8 \pi}{3}.
\label{gvalue}
\end{equation}
The construction \cite{zamzam} is based on four principles:

\noindent(i) Crossing symmetry.

\noindent(ii) Unitarity of the S-matrix.

\noindent(iii) Trilinear relation derived from the extra (infinite number of)
conservation laws associated with the sine-Gordon theory. 

\noindent(iv) Absence of the ``CDD pole'' ambiguity.

It is convenient to employ the rapidity variable $\theta_i$ for each pion
so that the momentum and energy become 
\begin{equation}
p_i=m_{\pi}\text{sinh}\; \theta_i, \quad E_i=m_{\pi}\text{cosh}\; \theta_i.
\end{equation}
The relevant variable is 
\begin{equation}
\theta \equiv \theta_1 - \theta_2
\end{equation}
in terms of which the Mandelstam variable reads
\begin{equation}
s=4m_{\pi}^2 \;\text{cosh}^2 (\theta/2).
\end{equation}
We see from (\ref{transmission}) that, for example, $S^{(+-)}_T$ depends
only on $s$, or equivalently on $\theta$.

The exact solution\footnote{We have complex conjugated (4.11) of
\cite{zamzam} in order that it reduce to their (4.12).} for the
soliton-anti-soliton transmission amplitude in the sine-Gordon model is:
\begin{eqnarray}
\lefteqn{S^{*}_T(\theta) =\prod_{l=0}^{\infty} 
\frac{\Gamma(\frac{l \gamma}{16 \pi}- \frac{i\theta}{2 \pi}) 
\Gamma(\frac{(l-1) \gamma}{16\pi} - \frac{i \theta}{2\pi})}
{\Gamma(\frac{1}{2} + \frac{l \gamma}{16 \pi}-\frac{i \theta}{2\pi})
\Gamma(-\frac{1}{2} + \frac{(l-1) \gamma}{16 \pi}-\frac{i \theta}{2 \pi})} 
\times } \nonumber \\
& &\times 
\frac{\Gamma(\frac{3}{2} + \frac{l \gamma}{16 \pi}+ \frac{i\theta}{2 \pi})  
\Gamma(\frac{1}{2} + \frac{(l-1) \gamma}{16 \pi}+ \frac{i \theta}{2 \pi})}
{\Gamma(1 + \frac{l \gamma}{16 \pi}+\frac{i \theta}{2\pi}) 
\Gamma(1 + \frac{(l-1) \gamma}{16 \pi}+\frac{i \theta}{2\pi})} .
\label{exacttransmission}
\end{eqnarray}
This rather complicated formula simplifies for the special case, as in
(\ref{gvalue}) when $\gamma = 8 \pi/n$, where $n$ is an integer. Using 
$\Gamma(z) \Gamma(1-z)= \pi/ \text{sin}(\pi z)$ we then get
\begin{equation}
S_T(\theta) = e^{in \pi} \prod_{k=1}^{n-1} \frac{e^{\theta -i(\pi
k/n)}+1}{e^{\theta}+e^{-i(\pi k/n)}}
\label{spcasetransmission}
\end{equation}
In the physical scattering region $\theta > 0$ this is just a pure phase
factor so there is no attenuation of the incoming wave. Furthermore, it is
easy to see that we also directly have $S_R(\theta)=0$, verifying that
there is no reflected wave. This special case was first discussed in
\cite{korfad}. 

Our model requires us to set $n=3$ in (\ref{spcasetransmission}). It is
amusing to note that the strong coupling limit of massive two-flavor
$QED_2$ is not merely an integrable model but one which corresponds to
reflectionless scattering. Even though (\ref{spcasetransmission}) is a pure
phase in the physical region, its general analytic structure is of
interest. The $n=3$ case has two poles in the unphysical region where
$\theta$ is pure imaginary. These are at:
\begin{eqnarray}
\theta=\frac{\pi i}{3}, (s=3m_{\pi}^2) &\text{where}& S_T(\theta)=\frac{2
        \sqrt{3}i}{\theta - i \pi/3} + \dots, \nonumber \\
\theta=\frac{2\pi i}{3}, (s=m_{\pi}^2) &\text{where}& S_T(\theta)=\frac{-2
        \sqrt{3}i}{\theta - i 2\pi/3} + \dots \label{poleandresidue}
\end{eqnarray}
and correspond respectively to the $\sigma$ and $\pi$ bound states. In fact
the prediction of the pole position of the exact scattering result is used
\cite{coleman2} to argue for the exactness of the DHN formula
(\ref{breathermass}). To transform (\ref{poleandresidue}) to the $s$-plane
it is sufficient to note that, near the poles, we may replace
\begin{equation}
\theta - \theta_0 = [s(\theta)-s(\theta_0)]/\left[\frac{ds}{d \theta}
\right]_{\theta=\theta_0}
\end{equation}
with $ds/d\theta = 2 m_{\pi}^2 \text{sinh} \theta$. Then the residues at
the bound state poles in the $s$-plane are
\begin{eqnarray}
\text{Res}[S_T, s=m_{\pi}^2]&=&6m_{\pi}^2, \label{exactresidue1} \\
\text{Res}[S_T, s=3m_{\pi}^2]&=&-6m_{\pi}^2, \label{exactresidue2}
\end{eqnarray}
Now let us compare these exact results with the tree level results we
obtained in (\ref{treeresidue1}, \ref{treeresidue2}). The residue at the
pion pole $\mp \sqrt{3} \pi m_{\pi}^2 \approx \mp 5.44 m_{\pi}^2$ agrees to
within ten percent if we adopt the lower sign. This is encouraging since it
again indicates that the tree level results may be close to the exact
ones. Of course, we cannot compare the magnitude\footnote{However, with the
same choice for the signs in (\ref{treeresidue1}, \ref{treeresidue2}), it
appears that the signs of the two residues in (\ref{exactresidue1},
\ref{exactresidue2}) should be the same. Usually a ``wrong sign'' residue
is associated with a ``ghost'' particle. However it is also possible, at
least in the scattering regime, for rescattering effects to change the
effective sign. An example is provided in the case of the $f_0(980)$
particle in $\pi \pi$ scattering, in Section IV A of \cite{hasasc}.} of the
residue at the sigma pole since it was introduced in an {\it ad hoc} way
and involves the undetermined factor $\gamma^2$.

Finally, we expect that, when one goes to higher orders in perturbation
theory, $S_T$ given in (\ref{transmission}) will exponentiate and $S_R$
given in (\ref{reflection}) will get cancelled. A possible hint of this may
be perceived in the large $s$ behavior shown in (\ref{larges})- $S_R$ is
seen to fall off very much faster with increasing $s$ than does $S_T$.
 
\section{Discussion}
We calculated the tree level analog $\pi-\pi$ scattering amplitude in the
strong coupling limit of massive two flavor $QED_2$. A characteristic
new feature, compared to the four dimensional case, is the presence of a
three point pion vertex. This comes from the WZW term and is allowed
because the two dimensional pion has positive $G$-parity. 

The resulting pion exchange contribution has the identical dependence on
kinematical variables (appropriately restricted) as the vector meson
exchange contribution in the theory based on a four dimensional effective
low energy Lagrangian for QCD. Since the analog $QED_2$ theory represents
an {\it exact} bosonization it seems that there is a sense in which the
``minimal'' QCD effective Lagrangian should include both pseudoscalars and
vectors. Of course, this does not exactly agree with the organization of
the chiral perturbation theory expansion
\cite{weinberg,gasleu1,gasleu2,meissner}. However in that approach, many of
the leading order ``counterterms'' are dominated \cite{dorava,egpr} by
vector meson exchange. For a tree level treatment, as suggested by the
$1/N_c$ expansion, the vectors are very important phenomenologically
\cite{hasasc}. In addition, when calculating the properties of
nucleons-as-solitons derived from the low energy Lagrangian, the presence
of vector mesons is crucial for a satisfactory understanding of the
``short-distance'' effects like neutron-proton mass splitting \cite{jjpsw},
``proton spin current'' \cite{jpsvw}, and heavy baryon hyperfine splitting
\cite{hqssw1,hqssw2,ssvw}. In any event, it seems worthwhile to further
contemplate the connection between the QCD effective Lagrangian and its
dimensionally reduced version.

In Section 4 we compared the tree level $\pi-\pi$ scattering in $QED_2$
with the known exact result in the sine-Gordon theory. It was pointed out
(though it is an elementary observation from existing results) that $\pi^+
\pi^-$ scattering in strong coupling $QED_2$ is not merely given by a known
analytic formula but is actually reflectionless. This, of course, can only
be approximated at tree level. However the model has two bound states so
its analytic structure is of great interest. The locations of the pion and
sigma poles satisfying $m_{\sigma}^2 = 3m_{\pi}^2$ have been well
documented in the literature. Here we pointed out that the residue at the
pion pole is quite well described (to about 10\% accuracy) by the tree
level calculation. Certainly it would be desirable in the future to extend
the perturbative tree level calculation to higher orders.

The triviality of the scattering in the two dimensional theory is clearly
different from the four dimensional QCD case. Another difference concerns
the question of spontaneous breakdown of chiral symmetry, which is
well-known to be a characteristic feature of the $QCD_4$ effective low
energy Lagrangian (when the quark mass terms are neglected). On the other
hand, the spontaneous breakdown of chiral symmetry is ruled out in the two
dimensional case, according to the Mermin-Wagner-Coleman theorem
\cite{merwag,coleman1}. One may wonder how this feature gets displayed at
the effective Lagrangian level, since it is not manifestly evident from the
bosonized action (\ref{action}). A heuristic way of understanding this was
discussed in \cite{delsch1,delsch2} using an old-fashioned linear sigma
model \cite{gellev} containing both $\pi$ and $\sigma$ fields. This model
is not an exact bosonization and does not faithfully mirror all the desired
properties of the two dimensional theory. Nevertheless, it contains a
manifest potential function which lets one conclude that the predicted
ratio $R=m_{\sigma}^2/m_{\pi}^2 =3$ corresponds to a theory which will {\it
not} be spontaneously broken when the parameter $m$ in (\ref{action}) is
set to zero. A study of the topography of the potential using the methods
of ``catastrophe theory'' suggests \cite{delsch2} that a generalized
spontaneous breakdown regime is related to the range $R>9$ which is
expected to hold in QCD. This type of analysis also seems like a promising
direction for future work.

\bigskip

\noindent{\bf Acknowledgments}
We are happy to thank Francesco Sannino for very helpful discussions. This
work was supported in part by the U.S. DOE Contract No DE-FG-02-ER40231.

\bibliography{scattering}
\bibliographystyle{prsty}

\end{document}